\pdfoutput=1
\documentclass[a4paper,11pt]{article}
\usepackage{amsmath}
\usepackage{amssymb}
\usepackage{graphicx}
\usepackage{xfrac}
\usepackage{soul}

\usepackage[utf8]{inputenc}

\usepackage[square,numbers,comma,sort&compress]{natbib}
\usepackage[colorlinks=true,citecolor=blue,linkcolor=purple,urlcolor=purple]{hyperref}
\usepackage{geometry}
\usepackage[usenames,dvipsnames,svgnames,table]{xcolor}
\usepackage{booktabs}
\usepackage{slashed}

\def\varabstract{ }
\def\varkeywords{ }
\def\vararxivnumber{ }
\def\vartitle{ }
\def\varsubtitle{ }
\renewcommand{\title}[1]{\gdef\vartitle{#1}}

\renewcommand{\abstract}[1]{\gdef\varabstract{#1}}
\newcommand{\keywords}[1]{\gdef\varkeywords{#1}}

\newtoks\authtoks
\renewcommand{\author}[2][]{%
	\authtoks=\expandafter{\the\authtoks#2$^{#1}$\ }%
}
\newtoks\affiltoks
\newcommand{\affiliation}[2][]{%
    \affiltoks=\expandafter{\the\affiltoks{\item[$^{#1}$]#2}}%
}
\newtoks\emailtoks\newcounter{emailcounter}%
\newcommand{\emailAdd}[1]{%
\ifnum\theemailcounter>0\emailtoks=\expandafter{\the\emailtoks, \typeemail{#1}}%
\else\emailtoks=\expandafter{\typeemail{#1}}%
\fi
\stepcounter{emailcounter}}
\newcommand{\typeemail}[1]{\href{mailto:#1}{\tt #1}}

\renewcommand\maketitle{
	\newgeometry{margin=2cm}
	\pagestyle{empty}\setcounter{page}{0}
	{\huge\flushleft\sffamily\bfseries\vartitle\\\Large\varsubtitle\par}
\vskip6ex
{\large\bfseries\raggedright\sffamily\the\authtoks\par}
\vskip2ex
\begin{list}{}{%
\setlength{\leftmargin}{0.28cm}%
\setlength{\labelsep}{0pt}%
\setlength{\itemsep}{-3pt}%
\setlength{\topsep}{-\parskip}}
\itshape\small%
\the\affiltoks
\end{list}
\vskip2ex
\noindent\hspace{0.28cm}\begin{minipage}[l]{.9\textwidth}
\begin{flushleft}
\textit{E-mail:} \the\emailtoks
\end{flushleft}
\end{minipage}
\vskip5ex
\noindent{\renewcommand\baselinestretch{.9}\textsc{Abstract:}}\ \varabstract
\vskip5ex
\if!\varkeywords!\else\noindent{\textsc{Keywords:}}\ \varkeywords \vskip2ex\fi
\if!\vararxivnumber!\else\noindent{\textsc{ArXiv ePrint:}} \href{http://arxiv.org/abs/\vararxivnumber}{\vararxivnumber}\vskip2ex\fi

\newpage
\restoregeometry
\pagestyle{plain}

\setcounter{footnote}{0}
}

\usepackage[square,numbers,comma,sort&compress]{natbib}

\definecolor{MS}{rgb}{0,0,1}

\usepackage{ifthen}

	\newcommand{\barlimc}[7]{
  \pgfmathparse{\mypos+0.3}
  \edef\mypos{\pgfmathresult}
		\node[left,scale=0.6] at (0,\mypos) {#1};
		\pgfmathparse{#3 > 5 ? 1 : 0}
		\ifthenelse{\pgfmathresult=1}{
			\fill[#2] ($(0,\mypos)+(0,-0.1)$) rectangle +(5,0.2);
			\fill[white] ($(0,\mypos)+(3.5,-0.1)$) rectangle +(0.3,0.2);
			\draw[decoration={zigzag},decorate,#2,very thick] (3.4,\mypos) to +(0.5,0);
			\node[left,scale=0.6] at (5,\mypos) {#3};
			}{
			\fill[#2] ($(0,\mypos)+(0,-0.1)$) rectangle +(#3,0.2);
			\node[left,scale=0.6] at (#3,\mypos) {#3};
		}
		\fill[#4] ($(0,\mypos)+(0,-0.1)$) rectangle +(#5,0.2);
		\node[left,scale=0.6] at (#5,\mypos) {#5};
		\fill[#6] ($(0,\mypos)+(0,-0.1)$) rectangle +(#7,0.2);
		\pgfmathparse{#7 <0.3 ? 1 : 0}
		\ifthenelse{\pgfmathresult=1}{
			\node[right,scale=0.6] at (0,\mypos) {#7};
		}{
		\node[left,scale=0.6] at (#7,\mypos) {#7};
	}
}

\title{Fitting in or odd one out? Pulls vs residual responses in $b\to s \ell^+\ell^-$}

\author[1]{Bernat Capdevila,}\emailAdd{bcapdevila@ifae.es}
\author[2,3]{Ursula Laa,}\emailAdd{ursula.laa@monash.edu}
\author[2]{and German Valencia}\emailAdd{german.valencia@monash.edu}

\affiliation[1]{Universitat Aut\`onoma de Barcelona, Institut de F\'isica d'Altes\\ Energies (IFAE), The Barcelona Institute of Science and Technology, Campus UAB, 08193 Bellaterra (Barcelona).}

\affiliation[2]{School of Physics and Astronomy, Monash University,
Melbourne VIC-3800}

\affiliation[3]{School of Econometrics and Business Statistics, Monash University,
Melbourne VIC-3800}

\abstract{
New results in processes with an underlying quark transition $b\to s \ell^+\ell^-$ have been recently reported by the LHCb and Belle II collaborations. In this note we show how the main implications of a handful of new measurements can be understood with the tools introduced in our recent paper, arXiv:1811.10793, without the need to redo the global fits. We find that the main impact of the new results, due to $R_K^{[1.1,6]}$ from LHCb,  is a decrease in $C_{10\mu}^{NP}$ with a reduced uncertainty. We validate this conclusion by presenting the result of a new global fit.
}


\begin{document}

\maketitle

  
\newpage

Processes with an underlying quark transition $b\to s \ell^+\ell^-$ have shown deviations from the SM for several years now. The most interesting ones from the point of view of establishing evidence for new physics are the ratios $R_K$ and $R_{K^\star}$ which test lepton universality. In addition, a collection of results for branching ratios and details of the angular distributions, such as $P_5^\prime$, are known to deviate from the SM coherently, resulting in global fits that consistently disfavor the SM by up to 5 standard deviations. In a recent paper \cite{Capdevila:2018jhy}, we have studied in detail one such six-dimensional global fit to the Wilson coefficients $C_{i\ell}$ ($i=7^{(')},9^{(')},10^{(')}$, $\ell=\mu$) \cite{Capdevila:2017bsm} (with values of observables updated to January 2019), introducing metrics that quantify the effect of the individual observables on the global fit. 

Some of the measurements included in the fit have since been updated and a few different, but related ones, presented. In this note we study the effect of these new results on the global fits with the framework developed in \cite{Capdevila:2018jhy}, and {\it without} performing a new global fit. With this exercise we illustrate the usefulness of the tools introduced in \cite{Capdevila:2018jhy} to assess the impact of a handful of new measurements on an existing global fit. We show that these methods suffice to understand the overall picture rapidly. The conclusions of this analysis are validated against the results of an updated global fit.

\section{New results  reported in 2019}\label{s:res}

We collect here the results that have appeared recently. Two of them correspond to two of the observables already included in the global fit studied in  \cite{Capdevila:2017bsm},  $R_K^{[1.1,6]}$ (LHCb) and ${\cal B}(B_s\to\mu^+\mu^-)$. 

\begin{itemize}

\item The LHCb collaboration announced a new measurement of $R_K$ \cite{Aaij:2019wad} using part of the full Run-2 data set. A combination of the former Run-1 measurement and the new value yields the following average 
\begin{eqnarray}
R_K^{[1.1,6]}=\frac{{\cal B}(B\to K\mu^+\mu^-)}{{\cal B}(B\to Ke^+ e^-)}=0.846^{+0.060+0.016}_{-0.054-0.014} ~~(0.745\pm 0.097).
\end{eqnarray}
In parenthesis we quote the previous result \cite{Aaij:2014ora} as was used in \cite{Capdevila:2018jhy}. The errors are combined in quadrature and the larger one is chosen for asymmetric cases following the practice used in the global fit \cite{Descotes-Genon:2015uva}.

As pointed out by the LHCb collaboration, the new central value moves closer to the standard model prediction but the error is reduced in such a way that there is little change in the significance of the result, $2.5\sigma$ instead of the previous $2.6\sigma$ below the SM. 
\item The Belle collaboration has new results for $R_{K^\star}$ combining the data from charged and neutral channels \cite{Abdesselam:2019wac}:
\begin{eqnarray}
R_{K^\star}^{[0.045,1.1]}&=&0.52^{+0.36}_{-0.26}\pm 0.05 ~~(0.66\pm 0.11) \nonumber \\
R_{K^\star}^{[1.1,6]}&=&0.96^{+0.45}_{-0.29}\pm 0.11 ~~(0.69\pm 0.12) \nonumber \\
R_{K^\star}^{[15,19]}&=&1.18^{+0.52}_{-0.32}\pm 0.10 
\end{eqnarray}
The first two of these measurements, $R_{K^\star}^{[0.045,1.1]}$ and $R_{K^\star}^{[1.1,6]}$, are similar to the corresponding LHCb measurements used in the 2017 fit (and quoted in parenthesis for comparison), but they differ in that the LHCb measured $R_{K^*}$ in the $B\to K^{0*}$ channel whereas Belle measured an isospin average of the neutral $B\to K^{0*}$ and charged $B\to K^{+*}$ channels. Additional results were given for $R_{K^\star}^{[0.1,8]}$ and $R_{K^\star}^{[0.045,]}$, but they will not be used due to difficulties in treating the theoretical errors.

\item A combination of the new ATLAS result for the branching ratio of the leptonic decay $B_s\to\mu^+\mu^-$ with previous CMS and LHCb numbers was carried out in Refs.~\cite{Arbey:2019duh}~and~\cite{Aebischer:2019mlg}, with similar results. We quote the first one for definiteness, and again list the previous result in parenthesis,
\begin{eqnarray}
10^{9}\times{\cal B}(B_s\to\mu^+\mu^-)=2.65^{+0.43}_{-0.39}~(3.0\pm 0.67).
\label{172v1}
\end{eqnarray}

These combinations are based on a composition of the experimental two-dimensional likelihoods without accounting for the asymmetries in parameter space caused by the fact that both ATLAS and LHCb have only been able to provide upper bounds on ${\cal B}(B^0\to\mu^+\mu^-)$.
The same combination but using a naive weighted average was performed by Ref.~\cite{Alguero:2019ptt} and leads to a rather different conclusion so we list it separately,
\begin{eqnarray}
10^{9}\times {\cal B}(B_s\to\mu^+\mu^-)=2.94\pm 0.43~~(3.0\pm 0.67).
\label{172v2}
\end{eqnarray}

\item Very recently, the Belle collaboration \cite{Abdesselam:2019lab} also released new measurements of $R_K$ using the full Belle data sample of 711 $\text{fb}^{-1}$, superseding their previous analysis of a 605 $\text{fb}^{-1}$ sample: 
\begin{eqnarray}
R_{K}^{[1,6]}&=&0.98^{+0.27}_{-0.23}\pm 0.06 \nonumber \\
R_{K}^{14.18<q^2}&=&1.11^{+0.29}_{-0.26}\pm 0.07 
\end{eqnarray}
We do not use the other $q^2$ ranges quoted by Belle following Ref.~\cite{Alguero:2019ptt}.
\end{itemize}

\section{Pull}\label{s:pull}
The Pull assesses the differences between the theory predictions $T$ for a given point in parameter space (e.g. the SM or the BF point) and the measurements (observed value $O$). The metric can be further quantified by noting that the sum of the squares of the Pull is a proxy for the total chi-squared of the BF in the absence of correlations, $\chi^2_{NC}$. In the global fit studied in \cite{Capdevila:2018jhy}, 
\begin{eqnarray}
\chi^2_{NC}=\sum_i {\rm Pull(BF)}_i^2=\sum_i\left(\frac{(T_{BF,i}-O_i)}{\sqrt{\Delta_{exp,i}^2+\Delta_{BF,i}^2}} \right)^2 =119.
\end{eqnarray}
Here $\Delta_{exp,i}$ is the experimental error of observable $i$ and $\Delta_{BF,i}$ the corresponding
theory error as evaluated at the BF point.
The contributions to this number from the two updated observables as used in \cite{Capdevila:2018jhy}, along with their new values are listed below. For the second one we give two results as per Eqs.~\ref{172v1} and~\ref{172v2} respectively.
\begin{eqnarray}
{\rm Pull(BF)}^2_{R_K^{[1.1,6]}}&=&0.13 \xrightarrow[]{\rm new} 1.11 \nonumber \\
{\rm Pull(BF)}^2_{{\cal B}(B_s\to\mu^+\mu^-)}&=&0.002 \xrightarrow[]{\rm new} 
\begin{cases}
0.66 & \mbox{\cite{Arbey:2019duh,Aebischer:2019mlg}} \\
0.03 & \mbox{ \cite{Alguero:2019ptt} }
\end{cases}
\end{eqnarray}

The changes in these two observables would increase the $\chi^2$ of the previous best fit point by 1.7 (taking the larger number for the second one), roughly  $0.07\sigma$ and thus a negligible effect.

Similarly, the Pulls for the new measurements (calculated with the old best fit) are presented in Table~\ref{t:frac}.
\begin{table}[htp]
\caption{Square of the Pull(BF) for new results not present in the previous global fit}
\begin{center}
\begin{tabular}{|l|c|c|c|c|c|}\hline
ID &$R_{K^\star}^{[0.045,1.1]}$ & $R_{K^\star}^{[1.1,6]}$  & $R_{K^\star}^{[15,19]}$  & $R_{K}^{[1,6]}$ &$ R_{K}^{14.18<q^2}$ \\ \hline
$\frac{(T_{BF,i}-O_i)^2}{\Delta_{exp,i}^2+\Delta_{BF,i}^2}  $ & 0.85 & 0.24 &0.45 & 0.52 &1.18 \\ \hline
\end{tabular}
\end{center}
\label{t:frac}
\end{table}%
These results show moderate Pull values\footnote{Recall that the largest Pull values for observables included in the fit were found to be larger than 2~\cite{Capdevila:2018jhy}.} indicating that the BF point found in the original fit still provides a good description of the new measurements, with the possible exception of $ R_{K}^{14.18<q^2}$, which we discuss below.

As a consequence we expect that the BF point in an updated fit will not change significantly, and the approximation established in~\cite{Capdevila:2018jhy} may be used to gain additional insights as demonstrated in the next section. This conclusion is corroborated by a new fit we present in the last section.

\section{Variation in the fit}

In  \cite{Capdevila:2018jhy} we 
established a framework for the discussion of single observables in
the context of the full fit which can  be directly extended to new measurements. 
The description is based on the uncertainty in the fit and consists of comparing  the predictions at the best fit point to those on 
the envelope of the six-dimensional one-sigma region.
We use the Hessian approximation  to describe this envelope as the set of twelve points
defined by the intersections of the $1\sigma$ ellipsoid with its principal axes.
The axes are determined through singular value decomposition of the Hessian matrix
at the global minimum of the $\chi^2$ function, and we thus refer to the points as the SVD points.
Because of correlations in the parameter space, the SVD directions do not correspond to single Wilson coefficients, the relation can be read off Table 2 of~\cite{Capdevila:2018jhy} and the SVD parameter points are listed in Table 9 of the same reference.
When correlations are ignored, there are twelve residual responses $\delta_i^{j\pm}$ per observable $i$, one for each SVD point ($j\pm$). They are calculated as
\begin{eqnarray}
\delta_i^{j\pm}&=&\frac{(T_i^{j\pm}-T_{BF,i})}{\sqrt{\Delta_{exp,i}^2+\Delta_{BF,i}^2}}.
\end{eqnarray}
While correlations were found to play an important role in evaluating single observables in the context of the fit, we do not include them in this analysis.
This is justified for two reasons: the observables of primary interest, $R_K^{[1.1,6]}$ and ${\cal B}(B_s\to\mu^+\mu^-)$,  were found not to be sensitive to correlation effects;\footnote{This is by construction for ${\cal B}(B_s\to\mu^+\mu^-)$.} 
in addition, including correlations is challenging in studies of single observables. 

To quantify the discussion we note that values of $\delta_i$ near one correspond to a variation of one standard deviation in observable $i$ as measured by the total uncorrelated error. Using this as a guideline, we find that $R_K^{[1.1,6]}$ (LHCb) and ${\cal B}(B_s\to\mu^+\mu^-)$ are the only new results that produce large residual responses, and never for directions $1^\pm$ or $2^\pm$.\footnote{These two directions were shown to be very tightly constrained through other observables \cite{Capdevila:2018jhy}, leaving little room for variations in their predictions.} The potentially interesting residual responses are collected in Table~\ref{t:deltas}, which shows the numbers previously obtained from the global fit (first column) as well as their corresponding updates (second column). Since the residual response does not depend on the central value of the measurement, both Eq.~\ref{172v1}~and~Eq.~\ref{172v2} give the same results in this case.

\begin{table}[htp]
\caption{Values of $\delta^{i\pm}$ for $R_K^{[1.1,6]}$ and ${\cal B}(B_s\to\mu^+\mu^-)$. The numbers obtained from the 2017 global fit are shown in the first column and their corresponding  updates in the second one.}
\begin{center}
\begin{tabular}{c||c|c||c|c}
ID & \multicolumn{2}{c||}{$R_K^{[1.1,6]}$}& \multicolumn{2}{c}{${\cal B}(B_s\to\mu^+\mu^-)$}  \\ \hline \hline
$\delta^{3+}$ &-0.87&-1.35&0.92&1.27\\
$\delta^{3-}$ &0.92&1.44&-0.83&-1.14\\ \hline
$\delta^{4+}$ &-1.53&-2.39&-0.72&-0.99\\
$\delta^{4-}$ &1.69&2.65&0.78&1.08\\ \hline
$\delta^{5+}$ &0.37&0.58&0.09&0.13\\
$\delta^{5-}$ &-0.23&-0.35&-0.09&-0.12\\ \hline
$\delta^{6+}$ &1.19&1.86&0.98&1.35\\
$\delta^{6-}$ &0.53&0.82&-0.88&-1.21\\
\end{tabular}
\end{center}
\label{t:deltas}
\end{table}%

Note here that the updated residual responses can be obtained from the previous ones by a simple rescaling because the measured value of the observable does not enter their  computation. The residual responses are only concerned with the variation in the prediction within $1\sigma$ of the BF point. 
The rescaling is due to the total uncorrelated error in the normalization. 
Since observables $R_K^{[1.1,6]}$ and ${\cal B}(B_s\to\mu^+\mu^-)$ have reduced experimental errors, there is an overall increase in their residual responses. 
For example  the total error in $R_K^{[1.1,6]}$ (LHCb) is dominated by the experimental error, and thus all its residual responses will be
increased by a factor of $\Delta_{exp}^{old} / \Delta_{exp}^{new} \approx 1.6$.\footnote{The situation is of course different in the
case of an entirely new observable (one that is not in the fit), as discussed for the Q observables in~\cite{Capdevila:2018jhy}.}

The fact that no large residual responses show up for any of the new measurements of $R_{K,K^\star}$ by Belle, follows mostly from their large experimental uncertainties. The case of $ R_{K}^{14.18<q^2}$, which has a large Pull, but small residual responses is interesting. Whereas it does not affect the global fit significantly (small residual response), it is clearly outside the fit uncertainty (large Pull). Note however, that it is in agreement with the SM which also falls outside the fit uncertainty.
This example illustrates how a combined study of Pulls and residual responses can be used to systematically assess whether a new measurement is statistically consistent with the rest of the set. In case of much larger Pulls than those observed for the measurements reported here this could be used as a criterion for excluding certain observables from the fit. 

On the other hand, the new measurements of $R_K^{[1.1,6]}$ and ${\cal B}(B_s\to\mu^+\mu^-)$ are expected to provide new constraints on the fit.
Table~\ref{t:deltas} shows that  $R_K^{[1.1,6]}$ had already been singled out as providing important constraints to directions $4^\pm,6^+$,  and to a lesser extent also to directions $3^\pm$ (just below the arbitrary cutoff of 1 used in \cite{Capdevila:2018jhy}). The table also 
shows that the new measurements result in notable increases in the residual response for all directions, and that values of $\delta_3^{\pm}$ are now also above the cutoff. ${\cal B}(B_s\to\mu^+\mu^-)$  had large (but below our cutoff) values of $\delta^{3\pm,4^\pm,6^\pm}$ and now exceeds the cutoff in all these cases.  We therefore expect the new result to reduce the fit uncertainties in these directions, especially along direction $4$.

In Table~\ref{t:dsq} we show the squares of the residual responses for $R_K^{[1.1,6]}$ and ${\cal B}(B_s\to\mu^+\mu^-)$ to quantify the importance of the new results in comparison to other observables entering the fit. The first row in each case reproduces the previous values and the updated numbers are shown in the second row. For comparison, the last row reproduces the value of this metric for the most important observable for each direction in the previous global fit \cite{Capdevila:2018jhy}, along with its corresponding ID. Since this metric is larger for 
the new measurements than for the previous most important observable, we conclude that the former will provide the dominant constraint in most directions. The notable exception being direction 5, which corresponds mostly to $C^\text{NP}_{9\mu}$. This is not surprising, as this direction was found to get similar cumulative constraints from multiple angular observables instead of being dominated by a single one. At first glance, direction  $6^-$ also appears to be an exception, but this is one example where correlations are expected to have large effects. In particular, observable 68 is highly sensitive to such effects and is not in the top five for direction $6^-$ when correlations are included~\cite{Capdevila:2018jhy}. For this reason, the updated measurement of ${\cal B}(B_s\to\mu^+\mu^-)$ is likely the dominant constraint on direction $6^-$ as well. 

\begin{table}[htp]
\caption{Values of $\delta_i^2$ for  $R_K^{[1.1,6]}$ (LHCb) and ${\cal B}(B_s\to\mu^+\mu^-)$. Their previous numbers are shown  in the first row and the updated ones in the second row. The largest  $\delta_i^2$ in the global fit along with its ID (in parenthesis) for each direction is shown in the last row for comparison. The relevant IDs are $P_2(B \to K^* \mu\mu) [6-8] $  LHCb (49)~\cite{Aaij:2015oid}, $P_2(B \to K^* \mu\mu) [15-19] $  LHCb (57)~\cite{Aaij:2015oid},  $Br(B^0  \to K^{0 *}\mu\mu) [15-19] $  LHCb (68)~\cite{Aaij:2016flj},  $R_K^{[1.1,6]}$ LHCb (98)~\cite{Aaij:2014ora} and $Br(B_s \to \mu\mu) $ (172)~\cite{Aaij:2017vad}. }
\begin{center}
\begin{tabular}{|c||c|c|c|c|c|c|c|c|} \hline
ID & $|\delta^{3+}|^2$ &$|\delta^{3-}|^2$ &$|\delta^{4+}|^2$ &$|\delta^{4-}|^2$ &$|\delta^{5+}|^2$ &$|\delta^{5-}|^2$ &$|\delta^{6+}|^2$ &$|\delta^{6-}|^2$  \\ \hline 
 98 &0.76&0.85&2.30&2.90&0.14&0.05&1.40&0.28 \\
&1.80&2.10&5.70&7.00&0.34&0.12&3.50&0.67 \\ \hline 
172&0.85&0.69&0.52&0.61&0.01&0.01&0.96&0.77\\
&1.60&1.30&0.98&1.20&0.02&0.01&1.80&1.50\\ \hline 
$|\delta|^2_{\rm max}$ &1.0 (68) & 0.9 (57) & 2.3 (98) & 2.9 (98) & 0.9 (57) & 0.6 (49) & 1.4 (98) & 2.0 (68) \\ \hline 
\end{tabular}
\end{center}
\label{t:dsq}
\end{table}%

\begin{figure}[h!]
\centering{\includegraphics[width=.3\textwidth]{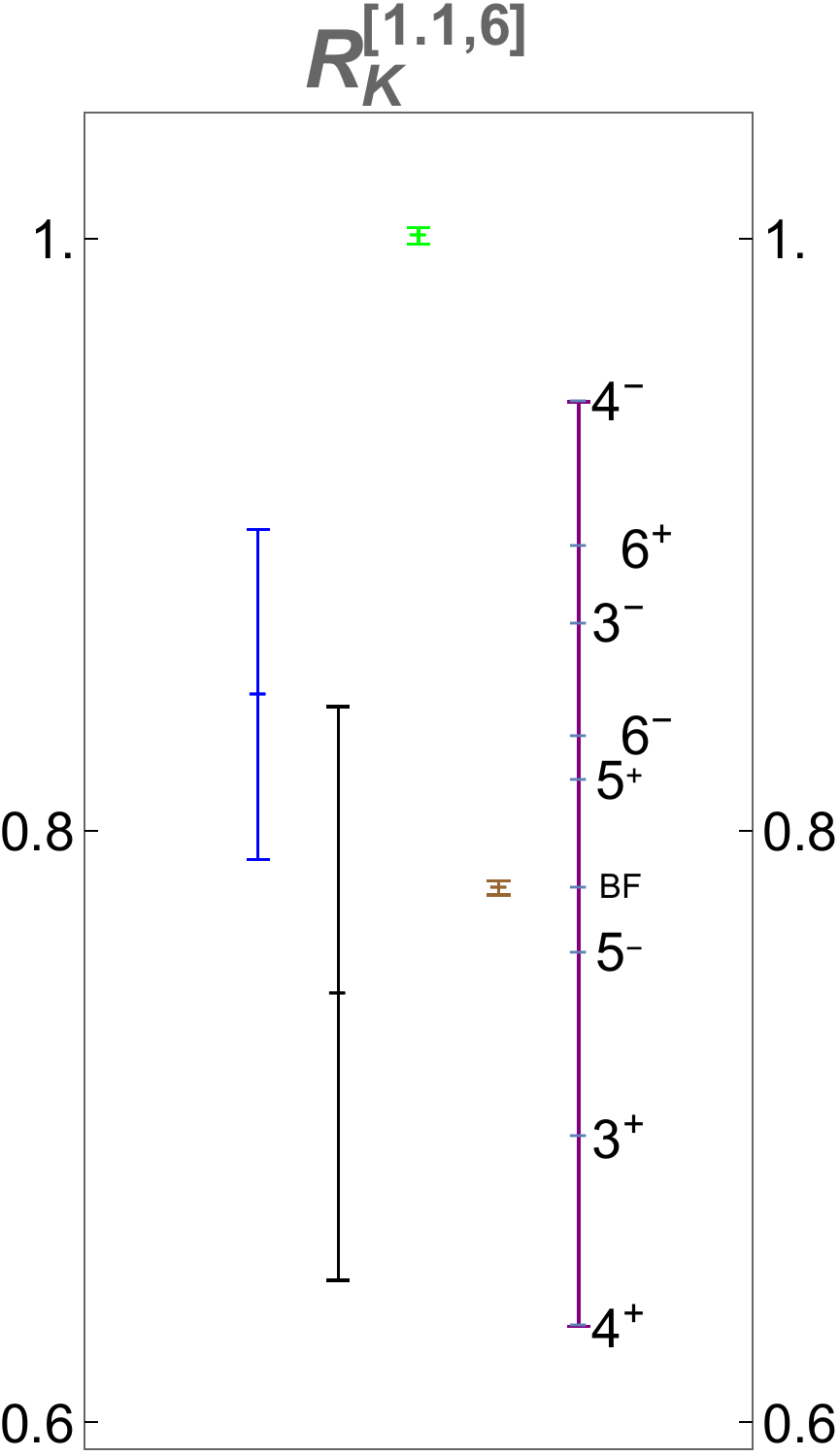}\includegraphics[width=.3\textwidth]{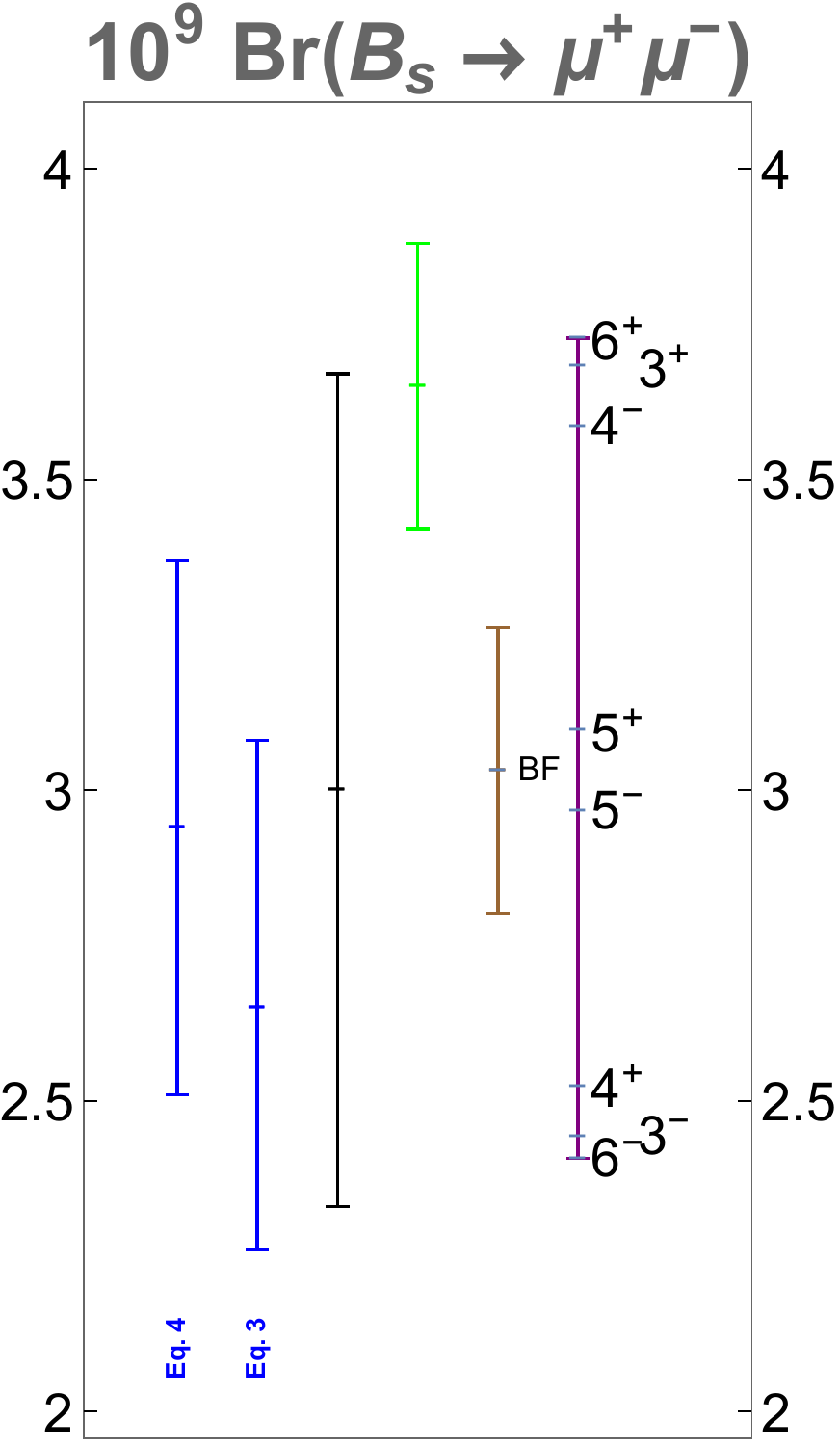}\includegraphics[width=.3\textwidth]{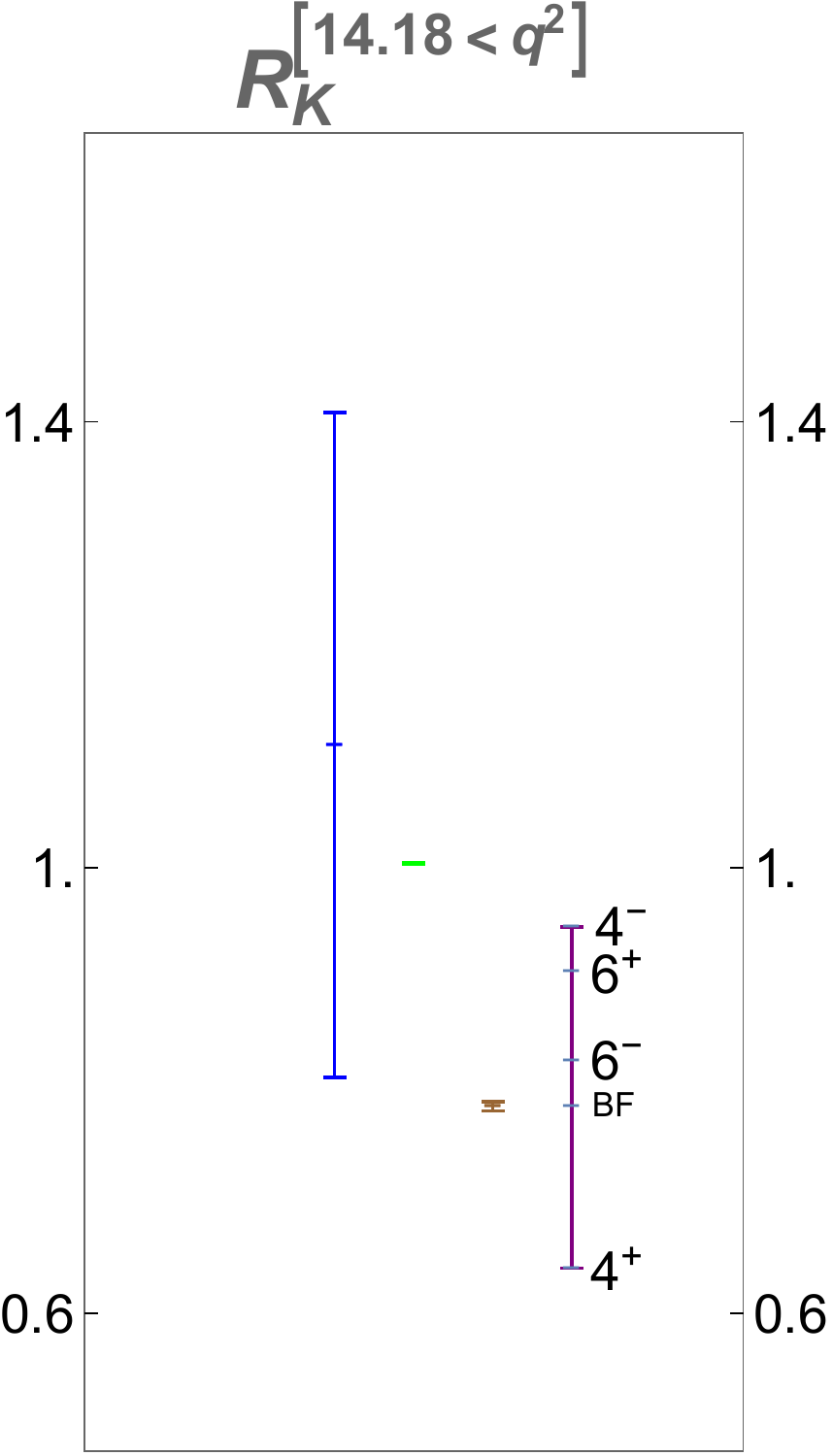}}
\caption{Comparison of errors for  observables $R_K^{[1.1,6]}$, ${\cal B}(B_s\to\mu^+\mu^-)$ and $ R_{K}^{14.18<q^2}$. In black (blue) the previous (new) measurement/average, in green the SM prediction, in brown the BF and in purple the fit uncertainty.}
\label{f:newd}
\end{figure}

In Figure ~\ref{f:newd} we compare the uncertainties for $R_K^{[1.1,6]}$, ${\cal B}(B_s\to\mu^+\mu^-)$ and $ R_{K}^{14.18<q^2}$. The black (blue) lines illustrate the previous and updated measurements respectively (there is no previous measurement for $ R_{K}^{14.18<q^2}$). These are to be compared with the SM prediction (green), the best fit (brown) and the fit uncertainty (purple). We have also labeled the position of the predictions at the SVD points that contribute the most to the fit uncertainty. The figure shows once more that the new measurements are consistent with the existing fit. In addition, it shows which directions in parameter space are preferred by the new measurements as explained below.

The full impact of the new measurements is obtained by studying Table~\ref{t:dsq} and Figure~\ref{f:newd} at the same time: the rankings indicate the importance of each observable in constraining the different directions in parameter space,
while the directional information in Figure~\ref{f:newd} indicates which of these directions are more consistent with the new results. 

\begin{itemize}

\item $R_K^{[1.1,6]}$ (LHCb) was already the most important observable for directions  $4^\pm,6^+$ as determined by the rankings. 
With the reduced experimental errors of the new measurement, it becomes the most important observable in directions $3^\pm$ as well, and completely dominates  $4^\pm$. 

Figure~\ref{f:newd} shows that the shift in the measurement is most easily accommodated by moving along direction $4^-$ but may also be explained by moving along directions $6^+,3^-$.\footnote{Direction $4$ has the largest residual response, implying that a given shift in the prediction will result in the minimal increase in the overall $\chi^2$ when it follows from changing the parameters along that direction. The required shift in this case, if taken along direction $3^-$ for example, requires a displacement of almost $1\sigma$  from the BF point.} 

Interestingly both $6^+$ and $6^-$ are on the same side of the BF point for this observable,  confirming the behaviour already identified in Fig. 9 of~\cite{Capdevila:2018jhy} now for the updated value of $R_K^{[1.1,6]}$. This is understood from theoretical expressions, which show
the relevant contributions linear in $C^\text{NP}_{10\mu}$, $C^\text{NP}_{9^\prime\mu}$ and $C^\text{NP}_{10^\prime\mu}$ approximately cancelling, with the result now showing a dominant quadratic dependence on the Wilson coefficients. This can be verified explicitly with the approximation given in Eq.~3 of \cite{Alguero:2019ptt}, which we find to be valid for the parameter points considered here.

The newly found dominance in the rankings of $R_K^{[1.1,6]}$ (LHCb) along direction 4 (mostly $C^\text{NP}_{10\mu}$ and to lesser extent $C^\text{NP}_{9\mu}$), places it as the determining observable for $4^\pm$, much in the same way that $B\to X_s\gamma$ dominates directions $1^\pm$.  We can also see that this new value of $R_K$ shifts the fit away from the previous BF in the direction of the SM along direction 4 towards $4^-$, hence implying a decrease in $C^\text{NP}_{10\mu}$. This specific observation is not possible  without this framework, and is corroborated by a new global fit as shown below.

\item A shift towards $4^-$ is also preferred by the new Belle measurement of $ R_{K}^{14.18<q^2}$.  Fig.~\ref{f:newd} illustrates why we find a large Pull but only small values of $\delta^{\pm}$ for this observable. It also shows that the measured central value cannot be accommodated within the $1\sigma$ region around the BF point. A similar situation was previously observed for the LHCb measurement of $ R_{K^\star}^{[0.045,1.1]}$~\cite{Aaij:2017vbb}, as discussed in~\cite{Capdevila:2018jhy}.

\item For ${\cal B}(B_s\to\mu^+\mu^-)$ the interpretation depends on the prescription used for updating the central value.
Following the result quoted in Eq.~\ref{172v1}, we see that the new central value suggests a shift towards $6^-$ and/or $3^-$,
whereas the central value quoted in Eq.~\ref{172v2} is aligned with the previous BF and results in a negligible Pull.

Note that Eq.~\ref{172v2} is used in the validation fit presented in Section~\ref{sec:val}, and therefore the updated results
for ${\cal B}(B_s\to\mu^+\mu^-)$ are expected to be reflected only in reduced fit uncertainty, without a shift in the BF point.
Using Eq.~\ref{172v1} instead, would likely result in an additional shift in the BF.  The rankings presented in Table~\ref{t:dsq} suggest in this case that the effect is much smaller than the one $R_K^{[1.1,6]}$ has on direction $4^-$.

\end{itemize}

\section{Validation}
\label{sec:val}

The purpose of this note is to show the usefulness of the tools introduced in Ref.~\cite{Capdevila:2018jhy} for assessing changes and/or additions to the observables in a global fit without the need to redo it. The main results found with this approach are that the updated measurement of $R_K^{[1.1,6]}$ is the most relevant for the global fit;  that we expect reduced fit uncertainties in most directions of parameter space; and that the BF point is changed along direction $4^-$  towards a smaller value of $C^\text{NP}_{10\mu}$.

To validate our results, we have redone the global fit including the new results detailed in Section~\ref{s:res}, obtaining a BF point with $\Delta\chi^2=39.1$ with respect to the SM, or $5.0\sigma$ away. This is to be compared with the previous BF point with $\Delta\chi^2=39.9$ with respect to the SM, or $5.1\sigma$ away. This change in $\Delta\chi^2$ validates our use of $\chi^2_{NC}$ as a proxy in Section~\ref{s:pull}.
The Wilson coefficients at the old/new BF points were/are:
\begin{eqnarray}
C^\text{NP}_7 = 0\xrightarrow[]{\rm new~BF}0.01,&& C^\text{NP}_{7^\prime}=0.02\xrightarrow[]{\rm new~BF}0.02, \nonumber \\
C^\text{NP}_{9\mu} = -1.06\xrightarrow[]{\rm new~BF}-1.13, && C^\text{NP}_{9^\prime\mu}=0.37\xrightarrow[]{\rm new~BF}0.44,  \nonumber \\
C^\text{NP}_{10\mu} = 0.34\xrightarrow[]{\rm new~BF}0.14, && C^\text{NP}_{10^\prime\mu}=-0.04\xrightarrow[]{\rm new~BF}-0.12\ .
\label{bfp}
\end{eqnarray}

The only significant change (when comparing to the profiled $1\sigma$ interval) is observed in $C^\text{NP}_{10\mu}$ which decreases in the new fit.

We can further test our predictions by considering the SVD basis instead, where the old/new BF points are
\begin{eqnarray}
v_1 = -0.030 \xrightarrow[]{\rm new~BF}-0.025,&& 
v_2=-0.016\xrightarrow[]{\rm new~BF}-0.018, \nonumber \\
v_3 = 0.011\xrightarrow[]{\rm new~BF}0, && 
v_4=0.71\xrightarrow[]{\rm new~BF}0.55,  \nonumber \\
v_5 = 0.87\xrightarrow[]{\rm new~BF}1.0, && 
v_6=0.35\xrightarrow[]{\rm new~BF}0.33\ ,
\label{bfp2}
\end{eqnarray}
with $v_i$ being the components of the BF point in the aforementioned basis. 
In this case we find that in addition to the expected shift along direction $4$ there is also some change along direction $5$, while movement along all other directions is negligible. The change along direction $5$ could not be expected from our analysis, and can be understood as follows. As pointed out in~\cite{Capdevila:2018jhy} there is no single measurement that is dominant in determining the BF point along this direction, which makes it more sensitive to small changes in the $\chi^2$ function that cannot be predicted in this framework.

\begin{figure}[h!]
\centering{\includegraphics[width=.45\textwidth]{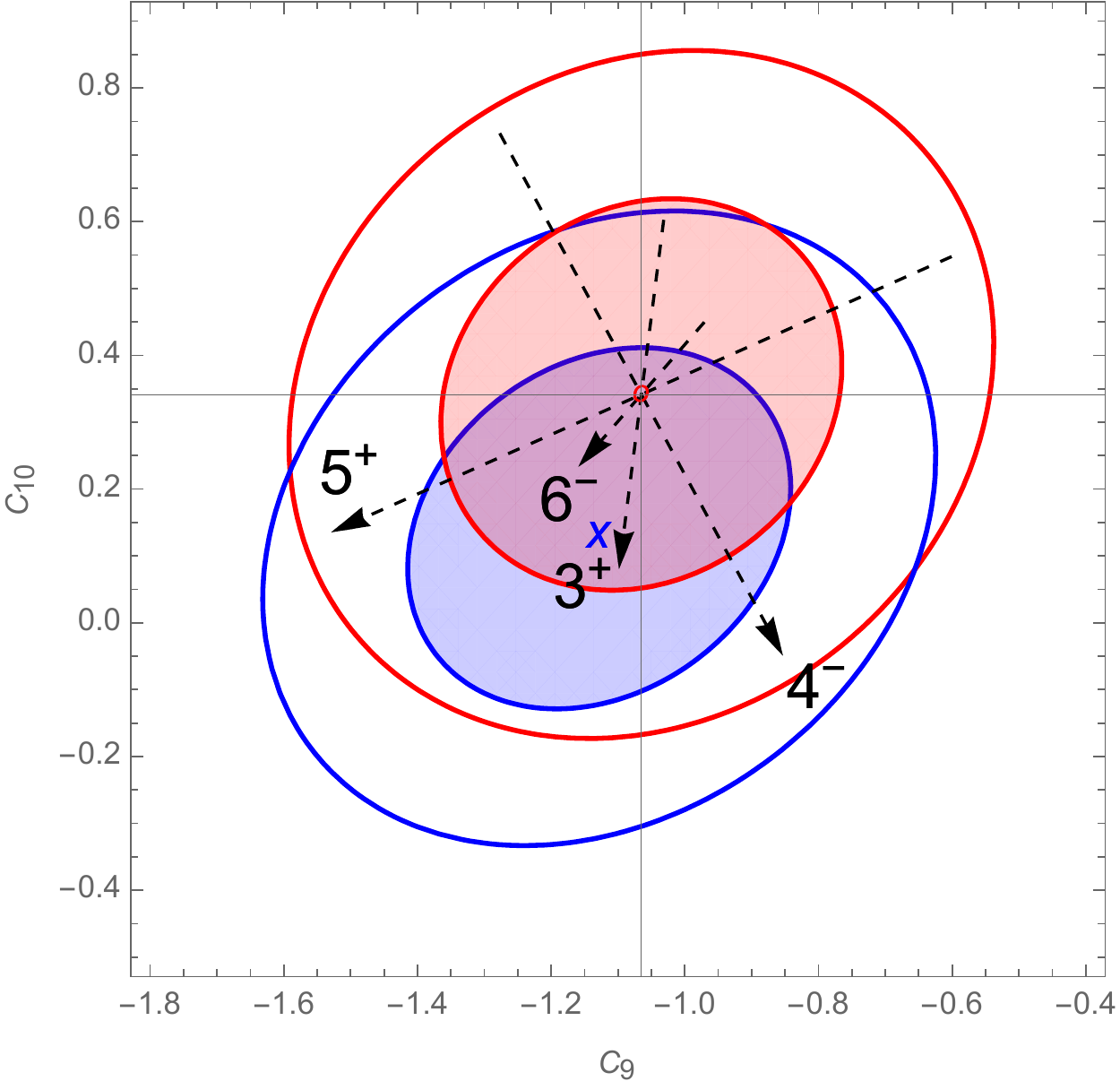}\includegraphics[width=.45\textwidth]{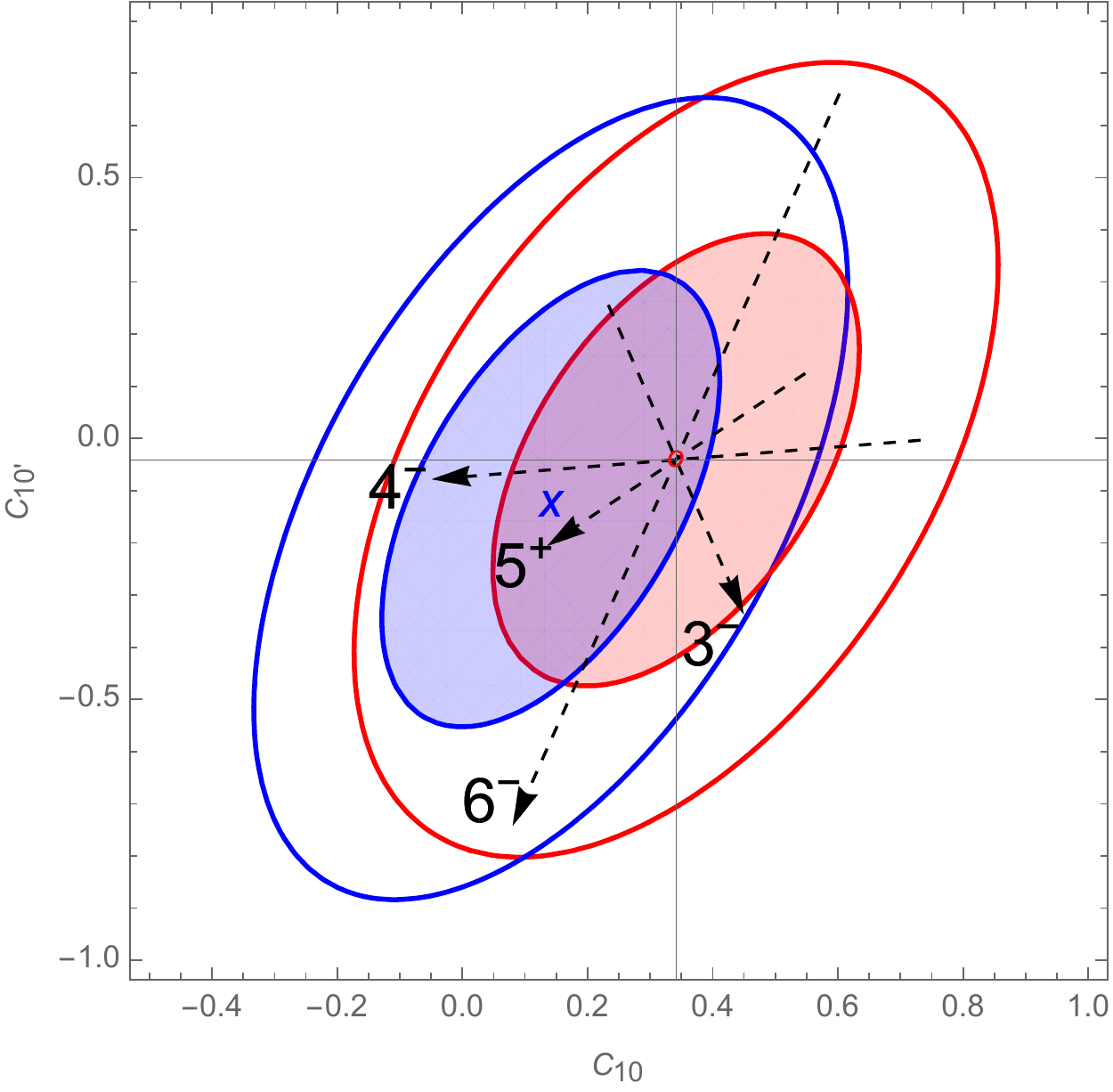}}
\caption{Comparing the old fit (blue) and new fit (red) in the $C^\text{NP}_{9\mu}-C^\text{NP}_{10\mu}$ and $C^\text{NP}_{10\mu}-C^\text{NP}_{10^\prime\mu}$ planes. The red $\circ$ (blue $\times$) marks the old (new) BF point.
The contours show the $1\sigma$ region obtained in the Hessian approximation where the full ellipse shows the profiled $1\sigma$ region while the outer ellipse is the projection of the full $1\sigma$ region.
The dashed lines show the projections of the most relevant SVD directions with the direction of the arrow indicating movement away from the old BF as suggested in the discussion.}
\label{f:other}
\end{figure}

Figure~\ref{f:other} shows a detailed comparison of the old and new fit the $C^\text{NP}_{9\mu}-C^\text{NP}_{10\mu}$ and $C^\text{NP}_{10\mu}-C^\text{NP}_{10^\prime\mu}$ planes.
Results from the old fit are shown in red, and the new fit is shown in blue. The red $\circ$ (blue $\times$) marks the old (new) BF point.
The contours show the $1\sigma$ region obtained in the Hessian approximation where the full ellipse shows the profiled $1\sigma$ region while the outer ellipse is the projection of the full $1\sigma$ region.
The dashed lines show the projections of the most relevant SVD directions with the direction of the arrow indicating movement away from the old BF as suggested in the discussion. Note that the length of each dashed line is $2\sigma$, as it connects the 2 SVD points in each direction.

As seen from Eq.~\ref{bfp2} the change in coefficients can be understood as a combination of shifts along $4^-$ and $5^+$.
In addition we confirm that the most relevant reduction in the uncertainty is found along direction $4$ resulting in increased correlation between $C^\text{NP}_{9\mu}$ and $C^\text{NP}_{10\mu}$.

\section{Conclusions}

We have studied the effect of measurements released after Moriond 2019 on the global fit of \cite{Capdevila:2017bsm} with the framework of \cite{Capdevila:2018jhy}. We have found that none of these results affects direction 5 significantly which means that $C^\text{NP}_{9\mu}$ is not affected. As this was the only coefficient required by the global fit to differ from the SM by a large amount, we conclude that the new measurements do not alter the observed evidence against the SM.

We have pinpointed the SVD directions most affected by these new results, finding they push the global fit towards $4^-$.
This corresponds to a decrease in $C_{10\mu}^{NP}$, which is corroborated with a new fit, Eq.~\ref{bfp}. 
In addition our study suggests reduced uncertainty primarily along direction $4$ as confirmed in Fig.~\ref{f:other}.

Our framework also predicts that the new value of ${\cal B}(B_s\to\mu^+\mu^-)$, when treated as in Eq.~\ref{172v1}, would result in a further small shift in $C_{10^\prime}$ towards larger negative values. The validation fit, however, treats this observable as in Eq.~\ref{172v2} so it does not contain this effect. Both treatments of this observable result in larger residual responses which contributes to the smaller overall uncertainty in the fit.

The example of $R_K^{[14.18<q^2]}$ from Belle, in which the Pull is large but the residual responses are small, suggests a systematic way to assess the consistency of specific measurements with the global fit that could be used as a criterion for exclusion.

In conclusion we have demonstrated how the formalism developed in \cite{Capdevila:2018jhy} can be used for a quick evaluation of the impact of new measurements without performing a new global fit.
We emphasise that it provides a useful framework to share the results of a multivariate fit not limited to the application presented here, and
we encourage groups performing multidimensional global fits to publish the Hessian approximation to enable more detailed follow-up studies without access to the full fit.
This is already common in the PDF community, and for example \cite{Wang:2018heo} published all residual responses enabling the study presented in \cite{Cook:2018mvr}.

\section*{Acknowledgments} This work was supported in part by the Australian Government through the Australian Research Council. BC acknowledges financial support from the grant FPA 2017-86989-P and Centro de Excelencia Severo Ochoa SEV-2012-0234.

\bibliography{biblio}

\end{document}